\documentclass[a4paper,showkeys,floatfix,aps,pre,longbibliography,superscriptaddress,reprint]{revtex4-2}
\usepackage{graphics,graphicx}
\usepackage{amsmath,amssymb}

\usepackage{graphics,graphicx}
\usepackage{dcolumn,bm}
\usepackage{psfrag}
\usepackage{xstring}
\usepackage{color}
\usepackage{booktabs}
\newcommand{\srm}
{\affiliation{Department of Physics, SRM University - AP, Amaravati,
 Andhra Pradesh - 522240, India}}

\newcommand{\srmcse}
{\affiliation{Department of Computer Science and Engineering, SRM University - AP, Amaravati, Andhra Pradesh - 522240, India}}

\begin{document}

\title{Efficient strategy for Parallel Minority Games}

\author{Ankith Reddy Vemula}

\email{ankithreddy\_vemula@srmap.edu.in}
\srmcse
\author{Soumyajyoti Biswas}

\email{soumyajyoti.b@srmap.edu.in}
\srm
\srmcse

\begin{abstract}
We study the parallel Minority Game, where a group of agents, each having two choices, try to independently decide on a strategy such that they stay on minority between their own two choices. However, there are multiple such groups of agents, and some of them have common choices. This overlap brings in additional competition among the agents making the variance reduction a complex optimization problem. We study multiple stochastic strategies and find that the most efficient strategy is the one where the agents have just the memory of their last visit to the their alternative choice. The resulting dynamics, even though giving the lowest population variance among the strategies studied here, end up in a frozen state. However, the frozen state does not bring the the variance to its lowest possible value; a situation qualitatively analogous to spin-glass systems.
\end{abstract}

\maketitle
%
%

\section{Introduction}

The Minority Game has been a paradigmatic example of repeated choice multiplayer game representing competition for limited resource allocation \cite{mg1,mg2,mg3}. Particularly, it was introduced as a variant of the El Farol bar problem \cite{elfarol}, and subsequently applied to stock markets and other two choice scenarios (see \cite{mg4,mg_rev1,mg_rev2} for reviews). The primary objective of the Minority Game is that multiple (odd number) players will decide to choose, independently and repeatedly, between two options, such that they end up in the minority group for most of the times. In other words, the individuals in the minority win a positive payoff in each step. In general, the pay-off is independent of the crowd size in the minority. 

Clearly, a complete random choice would lead to a Gaussian population distribution with the fluctuation proportional to $\sqrt{N}$, where $N$ is the total population. A long standing question has been to find a strategy that would allow a reduction in the population fluctuation between the two choices compared to the complete random choice. Since the pay-off is independent of the crowd size, the maximum utilization in the game comes from a near equal distribution of the population between the two choices. As was proposed in Ref. \cite{mg1}, agents with a memory of their last few choices can evolve towards a non-dictated, deterministic strategy that could lead to an emergent cooperation among them and it eventually results in a significant reduction of fluctuation between the population of the two choices compared to a purely random choice that is devoid of any such cooperation. A well developed literature exists, particularly around this phenomenon of emergent cooperation \cite{mg1,mg2}, phase transition in the model \cite{mg3}, resulting inequality \cite{ineq}, controlling collective behavior \cite{herd}, applications to stock markets \cite{stock}, multi-strategy variant \cite{grp}, and also attempts in experimental verification (see e.g., ref. \cite{expt})  among other important observations in this game (see \cite{mg_rev1,mg_rev2} for reviews).

A stochastic strategy \cite{sasi}, inspired from a multi-choice, multi-player game, the Kolkata Paise Restaurant (KPR) problem \cite{kpr1,kpr2} (see \cite{kpr3} for a general introduction), utilized the additional information of the crowd size in formulating the strategies of the individual players. The proposed stochastic strategy could reduce fluctuation to the minimum possible value very quickly (in $\log \log (N)$ time).  It was later shown that the strategy was rather robust, in the sense that even an approximate knowledge of the crowd-size in each step could lead to a quick ($\log (N)$) convergence of population to the minimum possible fluctuating state \cite{utl}. Indeed, the minimum possible fluctuation is a potential drawback, since it is a frozen state for the stochastic dynamics: two choices with populations $m$ and $m+1$, where $N=2m+1$. A winning (minority) group would continue to remain in the minority, once this division is obtained. Escape from this frozen (absorbing) state required addition of random (irrational) agents. 

\begin{figure}
\includegraphics[width=8.5cm]{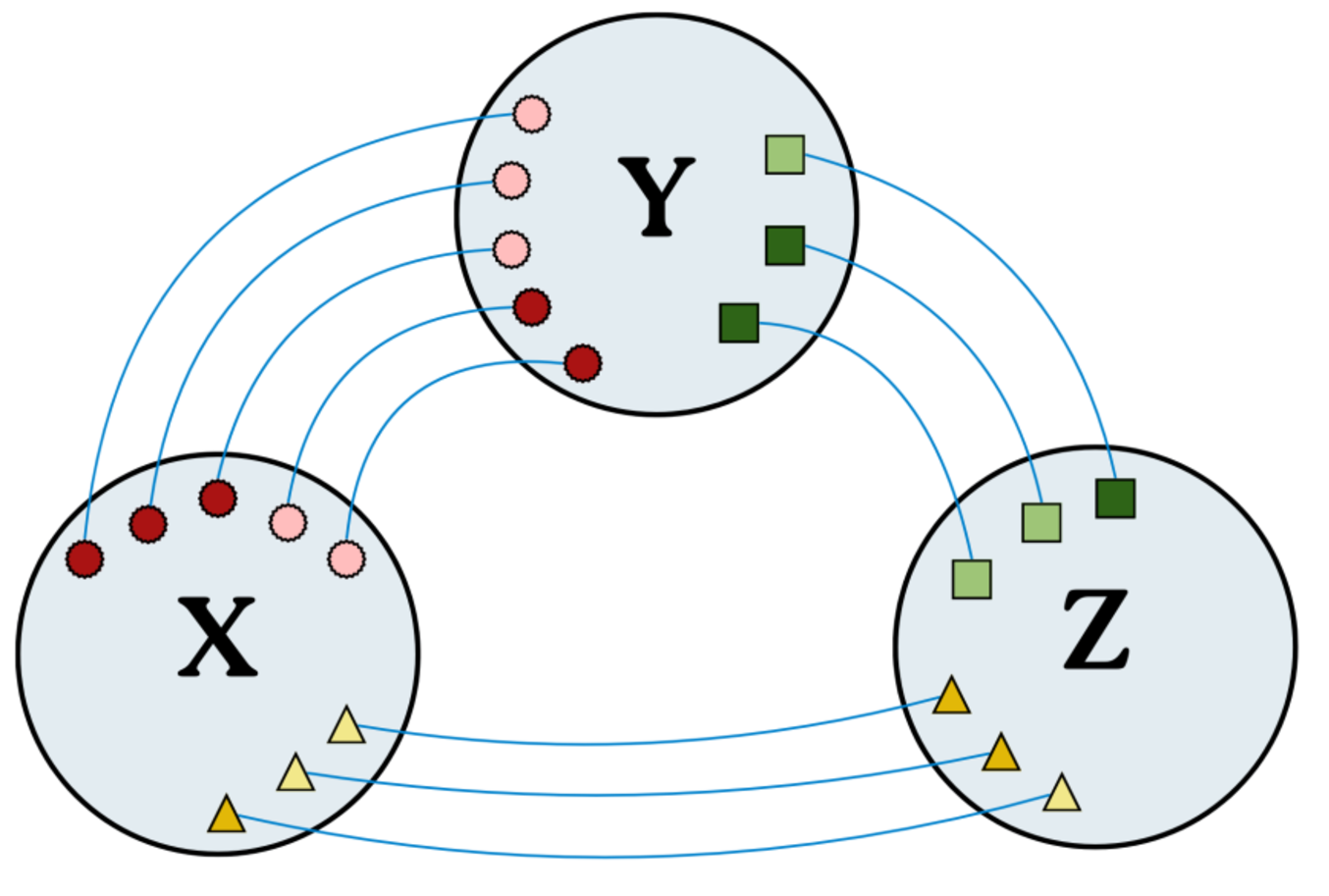}
\caption{A schematic representation of the parallel Minority Game is shown, with $D=3$, $N=11$.
A filled shape represents an occupied place and an empty shape represents a vacant place for the corresponding agent in their alternate location. Hence, the populations at the three locations are given by $n_X=4$, $n_Y=4$ and $n_Z=3$. As an example, the red-circles behave an a Minority Game between the choices $X$ and $Y$, so do the green-squares between $Y$ and $Z$ and so on. But for a given type of agents, say red-circles, there are other agents (green-squares and yellow-triangles) in both of their choices who affect whether a given red-circle is in the Minority or not. In the simulation results we take $D=101$ and vary $g=N/D$ between 21 and 151.}
\label{fig_1}
\end{figure}

In many realistic settings, the set of available alternatives may be large, say of size $D$, even though each agent is restricted to a fixed pair among them. Thus, an agent may only switch between its designated two choices. From the perspective of that agent, the resulting dynamics are analogous to those of the MG; however, the populations at its two choices are influenced by other agents whose respective alternate options are selected from the remaining 
$D-2$ choices with equal probability. 
This is the parallel minority game (PMG), introduced in ref. \cite{pmg}, which is a multi-player and restrictive multi-choice game. It is a set of minority games, played in parallel, where choices and players may overlap (see Fig. \ref{fig_1}). Such overlaps would result in an increase in fluctuation and thereby making overall fluctuation reduction much more challenging than in the MG or even the KPR, where the agents are free to choose among any choices. PMG was introduced in the context of population movement during the COVID-19 pandemic. However, this can be viewed in a more general context of MG with non-conserved population, and could be applicable to multi-stock portfolio and the individual buy-sell decision in each stock, rather than the single stock buy-sell in the MG.  

Here we study the effects of various stochastic strategies, drawn from some of the strategies originally proposed for the KPR problem. We find that due to the parallel nature of the game, the maximum possible utilization (minimum fluctuation) state is never reached. However, the best utilization is reached for a strategy that has a convergence time that grows in a power-law with the total population, unlike the fast convergence ($\log (N)$ or $\log\log (N)$) times noted for stochastic strategy MG. 

\section{Model and strategies}

The parallel Minority Game (PMG) \cite{pmg} consists of $N$ agents and $D$ choices. Each agent has just two choices, just like the MG. This means, at any instance, the agents present in any one of the $D$ choices have their other choice distributed randomly and uniformly among the remaining $D-1$ choices. This implies, any two choices make a classical Minority Game system, but the population on both of these choices are not fixed. In that sense, the total system can be viewed as a set of $D/2$ coupled Minority Games running in parallel. 

\begin{figure*}
\includegraphics[width=5.5cm]{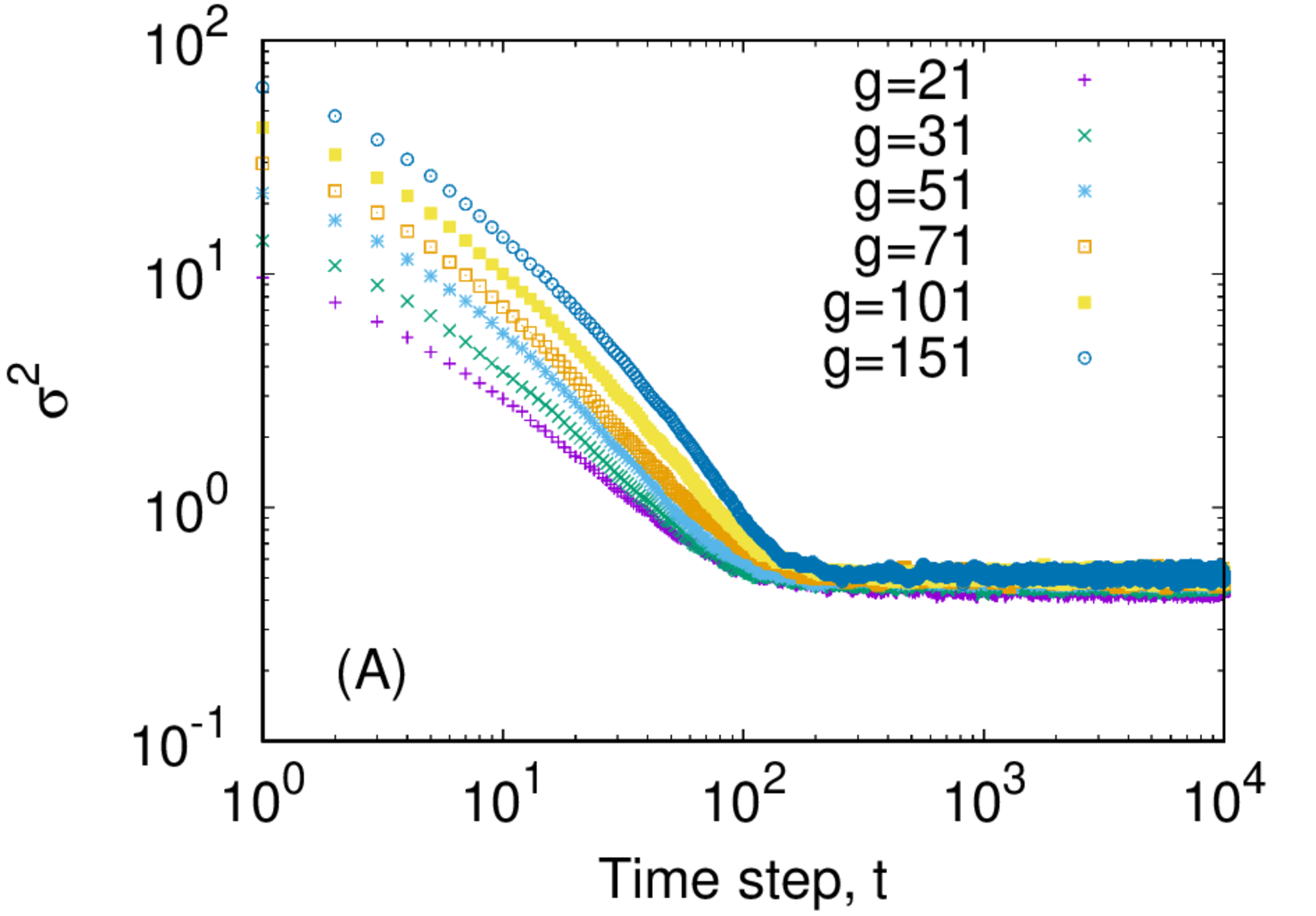}
\includegraphics[width=5.5cm]{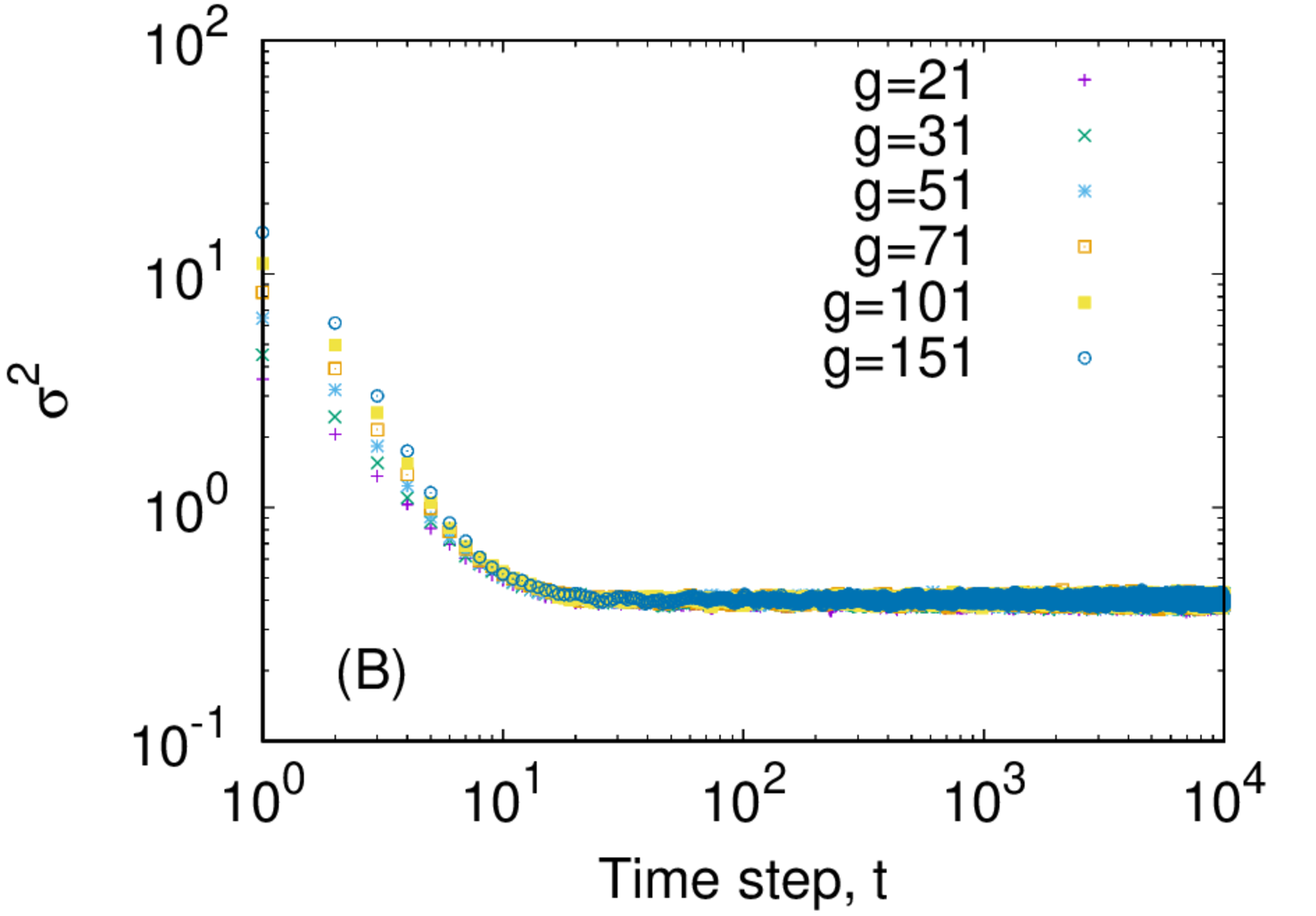}
\includegraphics[width=5.5cm]{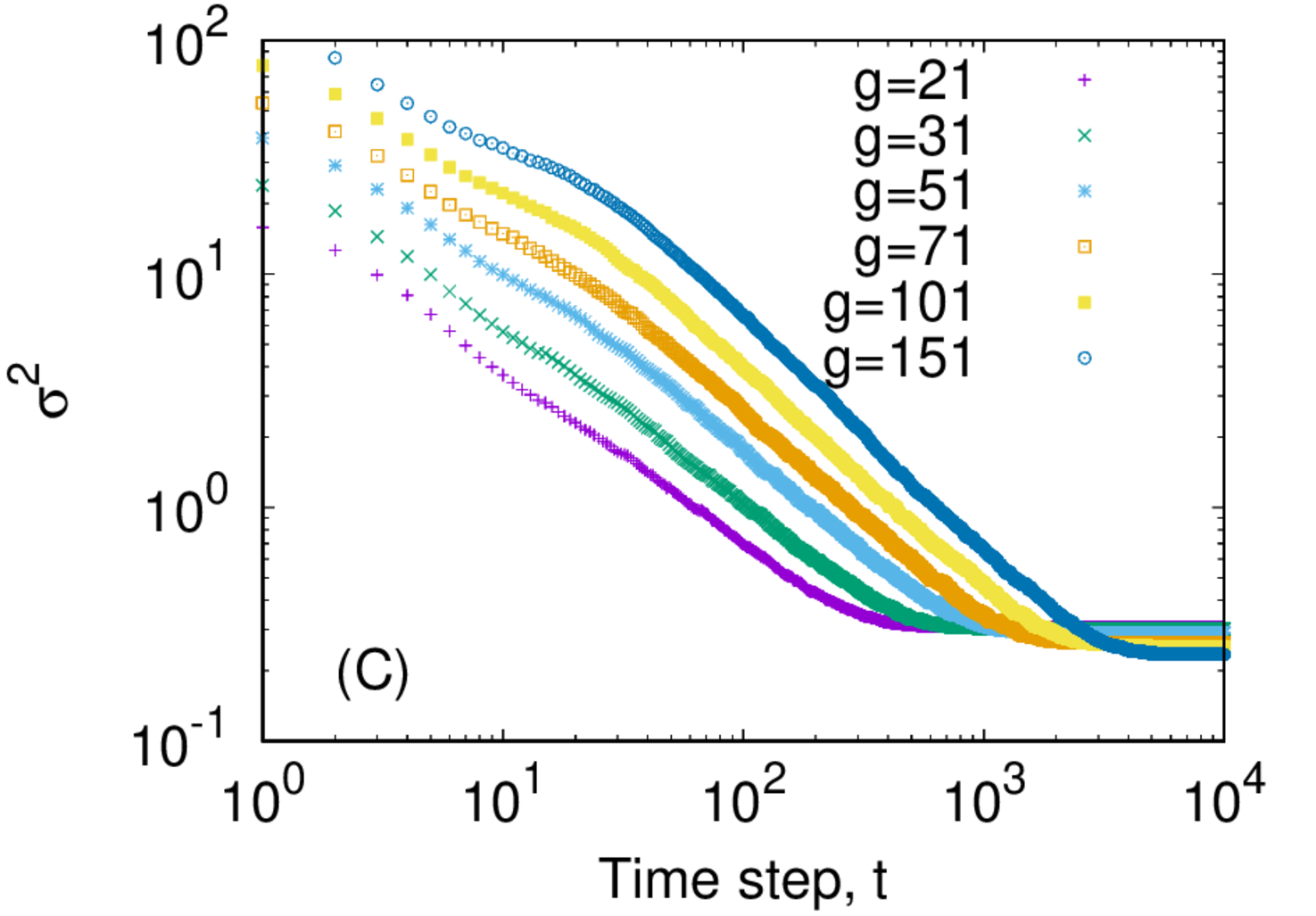}
\caption{The variance of population $\sigma^2$ in all $D$ choices are plotted for different values of $g=N/D$ (see Eq. (\ref{eq_rms})). The three figures correspond to the strategies A, B and C respectively. For the first two strategies, the saturation in variance comes rather fast and are almost independent of the system size $N=gD$, while for strategy C the variance decays in a power law and taken longer to saturate. Among these, the lowest value is obtained for the strategy C. However, as discussed later, the saturation in strategy C corresponds to a freezing of the dynamics. It is worth noting that for complete random choice, one would expect $\sigma^2\sim g$.}
\label{fig_2}
\end{figure*}

\begin{figure*}
\includegraphics[width=5.5cm]{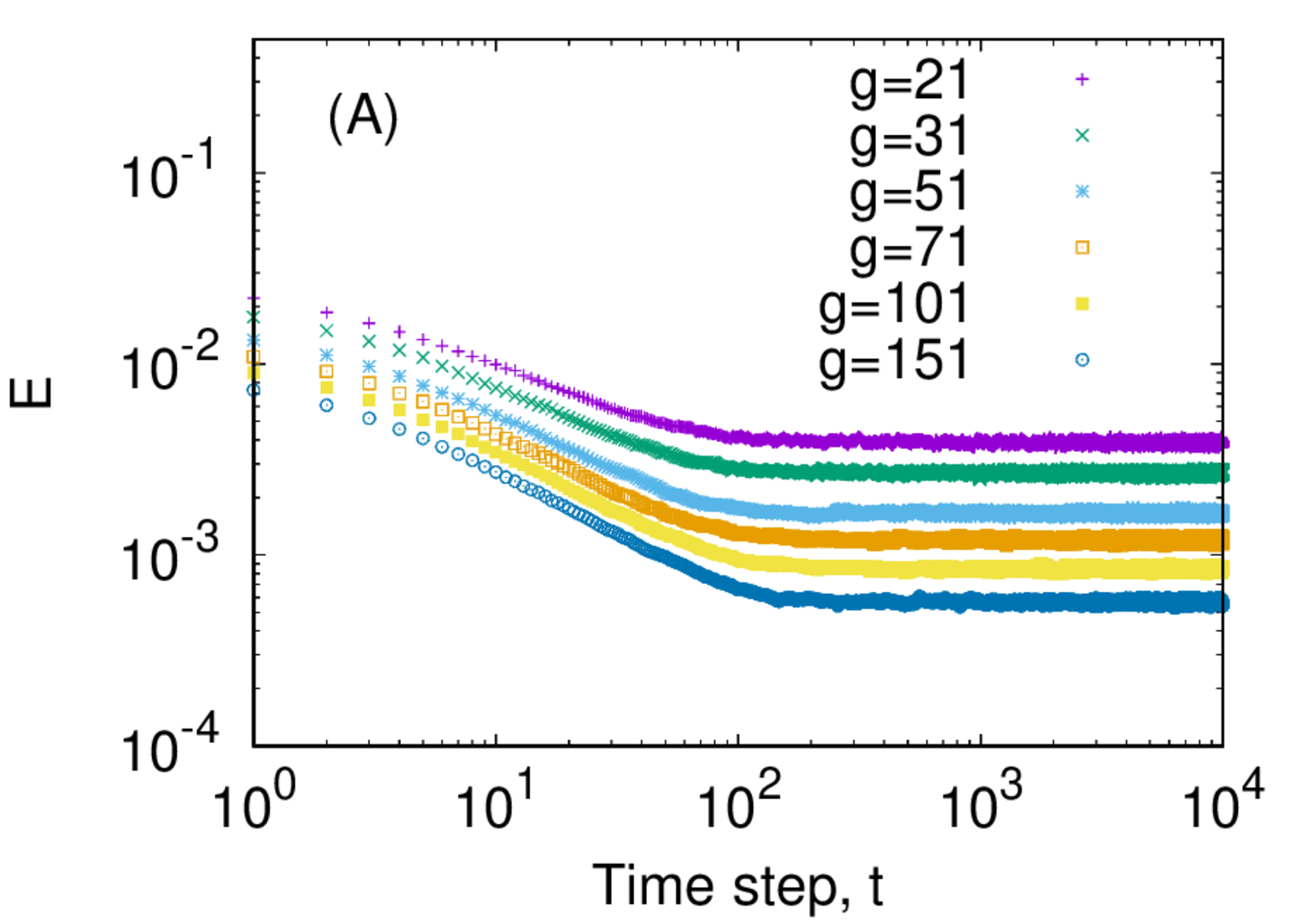}
\includegraphics[width=5.5cm]{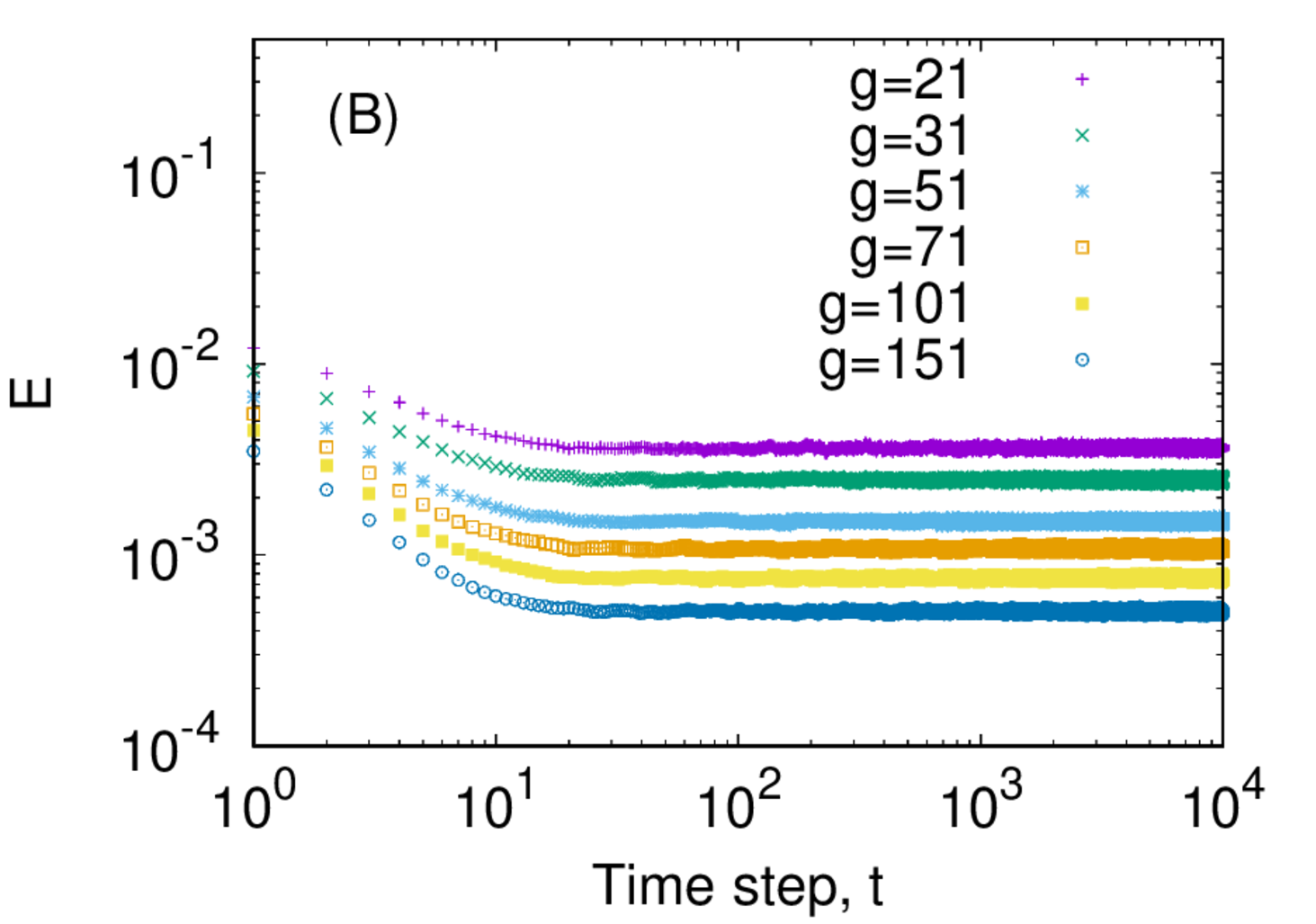}
\includegraphics[width=5.5cm]{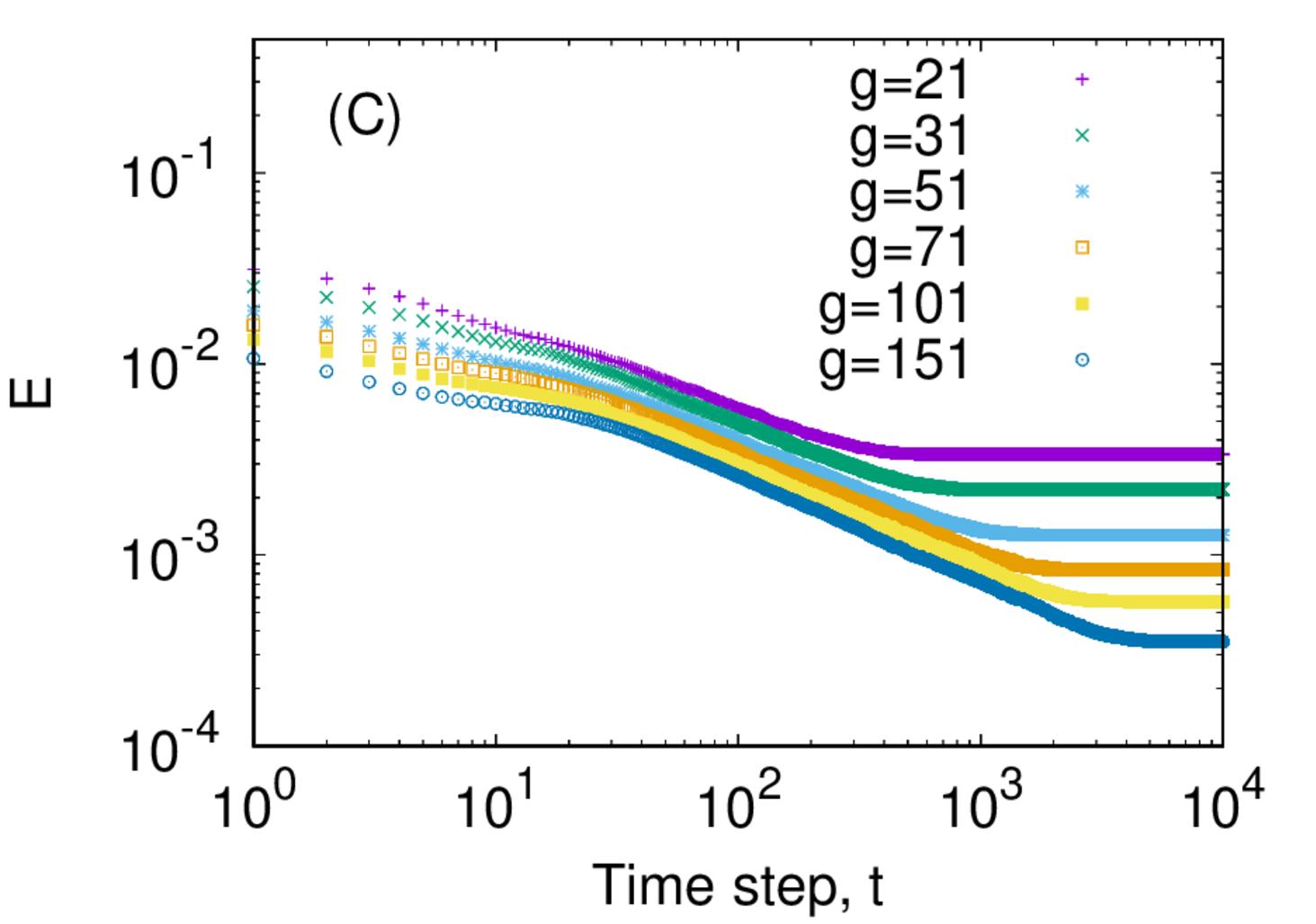}
\caption{The fraction of excess population $E$ (Eq. (\ref{eq_E})) is plotted for different values of $g$. The three figures correspond to the three strategies respectively. In all cases $E$ decreases with increase in $g$, i.e. system size. As before, the lowest values are obtained for strategy C.}
\label{fig_3}
\end{figure*}

The objective, as with the classical MG, remains that each agent through repetitive independent choices attempts to be in the minority location between their own two choices. The overall success of any (non-dictated) strategy is the population volatility ($\sigma$) in all the $D$ choices: 
\begin{equation}
\sigma^2(t)=\frac{1}{D}\sum\limits_{k=1}^{D}(n_k(t)-\langle n\rangle)^2, 
\label{eq_rms}
\end{equation}
where $n_k(t)$ is the population at the $k$-th choice at time $t$. A time step is defined as $N$ attempted switch by the agents, and the volatility is generally a function of that. Other than the volatility, keeping in mind that the populations in different locations may vary and therefore one agent can simply be in majority because their alternate location is over crowded, we define two other measures for excess populations. We measure the absolute number of agent whose alternate location is less crowded than their present location
\begin{equation}
e(t)=\frac{1}{D}\sum\limits_{i=1}^N \Theta(n_{x_i}(t)-n_{y_i}(t)),
\label{eq_e}
\end{equation}
where $\Theta(x)$ is the Heaviside function, with $\Theta(0)=0$; the current location for the $i$-th agent is $x_i$ and the alternate location of the same agent is $y_i$. As noted before, the location of the $i$-th agent can only take these two values. In that sense there is an explicit time dependence for the locations too, which we are not putting in the equation above for simplicity. We also measure the fraction of excess crowd as follows:
\begin{equation}
E(t)=\sum\limits_{i=1}^N\frac{n_{x_i}(t)-n_{y_i}(t)}{n_{x_i}(t)+n_{y_i}(t)}\Theta(n_{x_i}(t)-n_{y_i}(t)).
\label{eq_E}
\end{equation}
All of the above three measures quantify the population fluctuations, and are expected to tend towards zero when the population distribution is (nearly) uniform across all the available choices. 

Apart from the population fluctuations, there is another potential source of misutilization of resources, which is the duration up to which an agent stays in a given location. In particular, the stochastic strategies for MG \cite{sasi,utl} can lead to a situation where the switching dynamics stop and as a consequence the agents are stuck in their choices. This is, even though a Nash equilibrium, often unrelatable to real world applications. For example, if the choices were to indicate a buy/sell option for a stock \cite{stock}, then surely sticking to just one option would lead to an instability in the market at large and for those stuck agents in particular. It is, therefore, important to measure the residence time for an agent in a choice. To do that we map the switching dynamics to a virtual random walk (see e.g., \cite{walk} for similar such mapping from two-state systems). We assign a walker to each agent, such that the walker takes a step towards right at each instance the agent stays at their first location (say, at $x_i$) and takes a step towards the left at each instance the agent stays at their second location (say, at $y_i$).
Therefore, for this virtual walker, the total distance traveled will vary as
\begin{equation}
d_i(t+1)=d_i(t)+\xi_i(t),
\label{map}
\end{equation}
where $\xi_i=+1$ if the $i$-th agent is at $x_i$, and $\xi_i=-1$ if the agent is at $y_i$.
An agent stuck at a location would lead to a ballistic walker, whereas a regular switching would lead to a diffusive walker. The time dependence of the average distance traveled by the walker $\langle d(t)\rangle$ would accordingly be proportional to $t$ and $t^{1/2}$.


\begin{figure*}
\includegraphics[width=5.5cm]{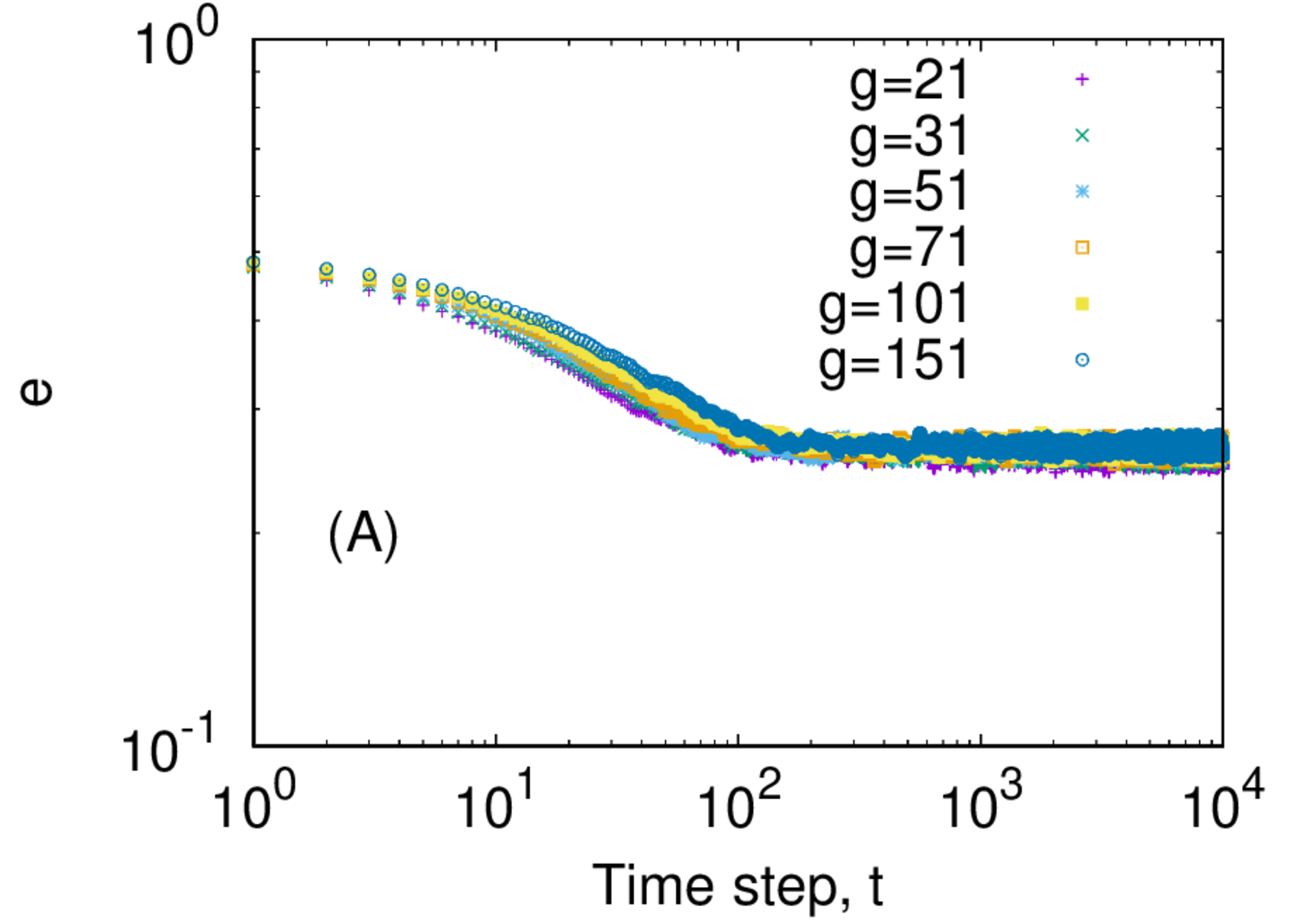}
\includegraphics[width=5.5cm]{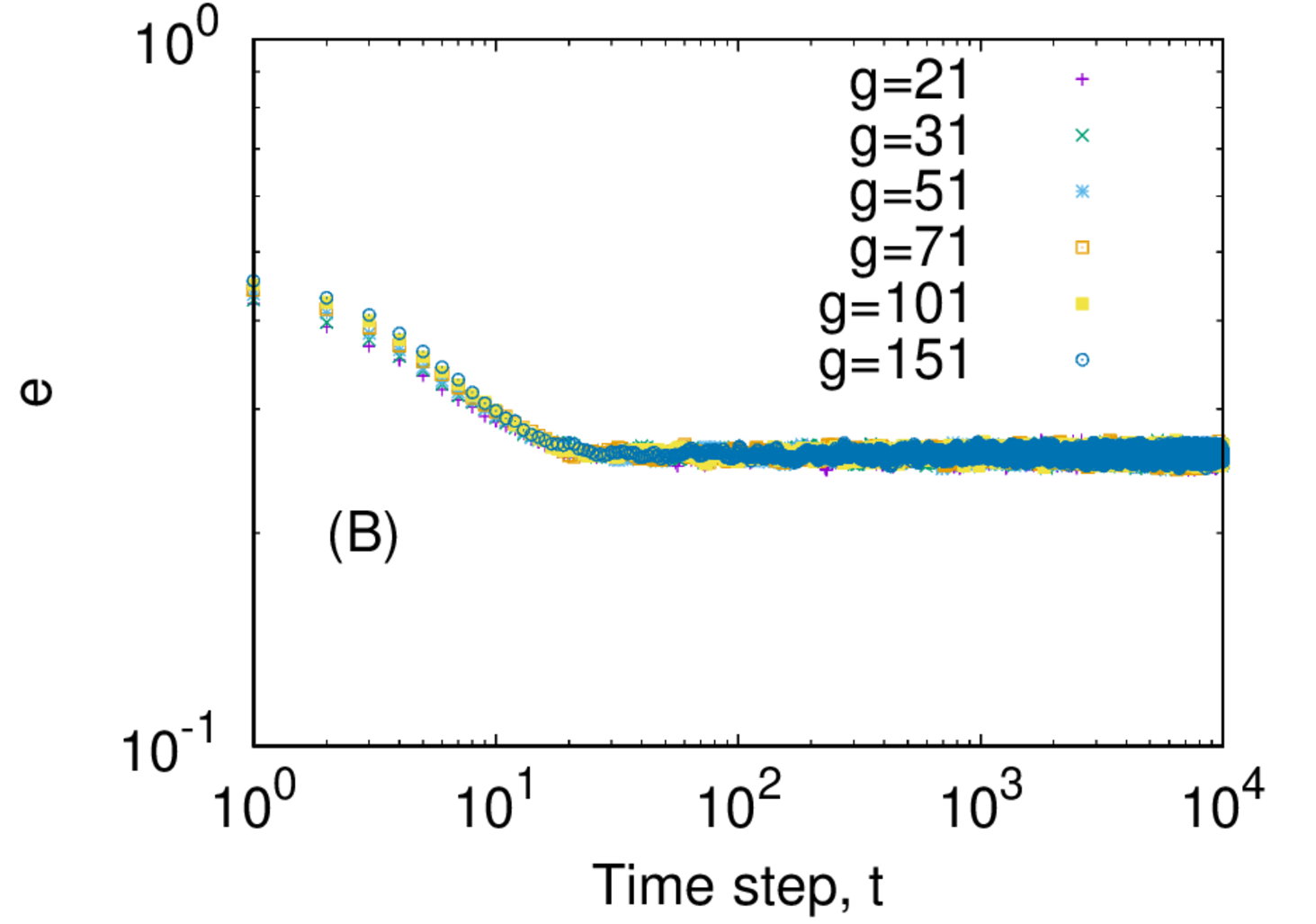}
\includegraphics[width=5.5cm]{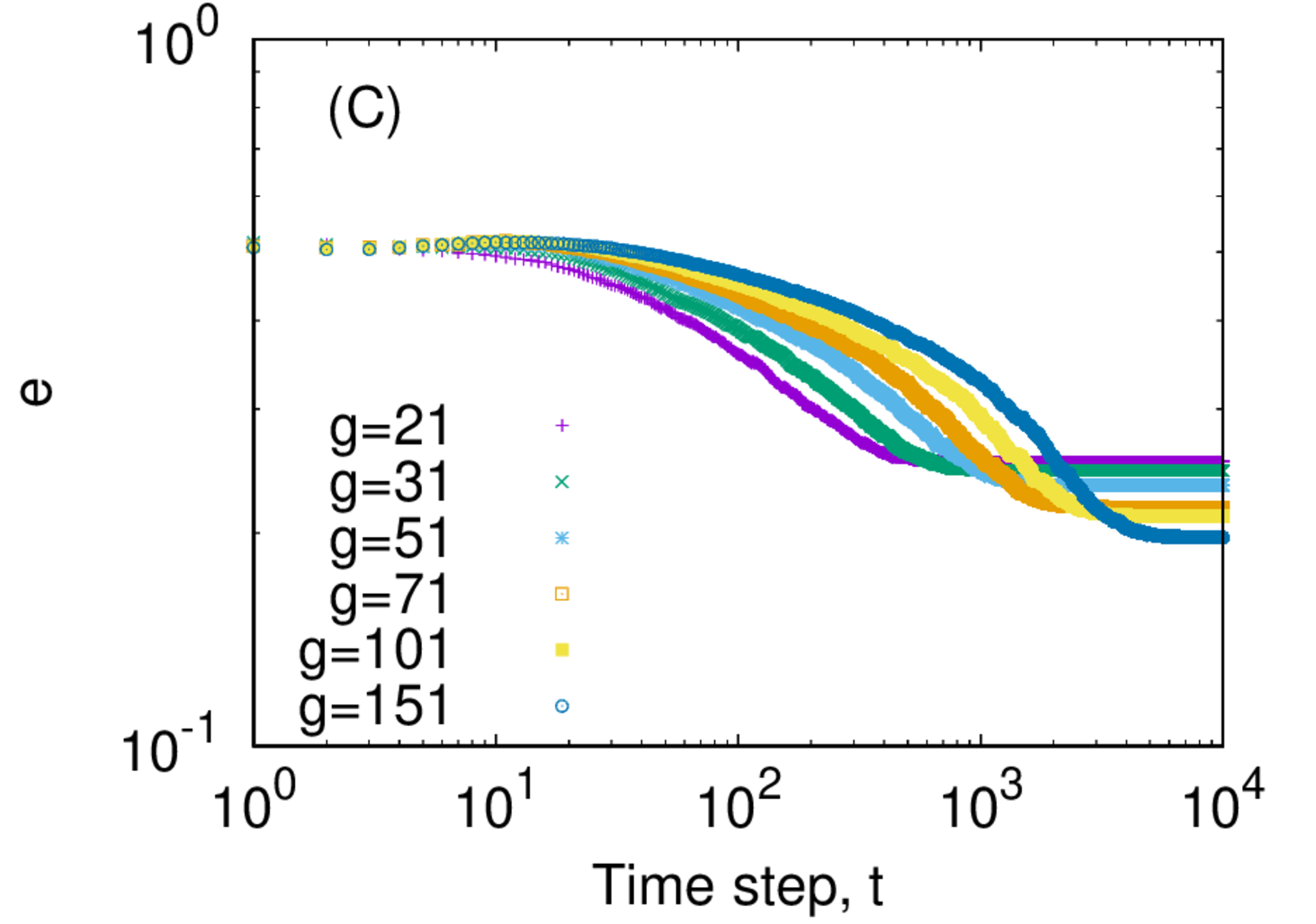}
\caption{The variation of the average number of agents in majority choices $e$ (Eq. (\ref{eq_e})) is shown for different values of $g$ for the three strategies. There is no perceivable difference in the saturation values for the first two strategies, while for strategy C it decreases with $g$. The scaling properties are discussed later on.}
\label{fig_4}
\end{figure*}

\begin{figure*}
\includegraphics[width=5.5cm]{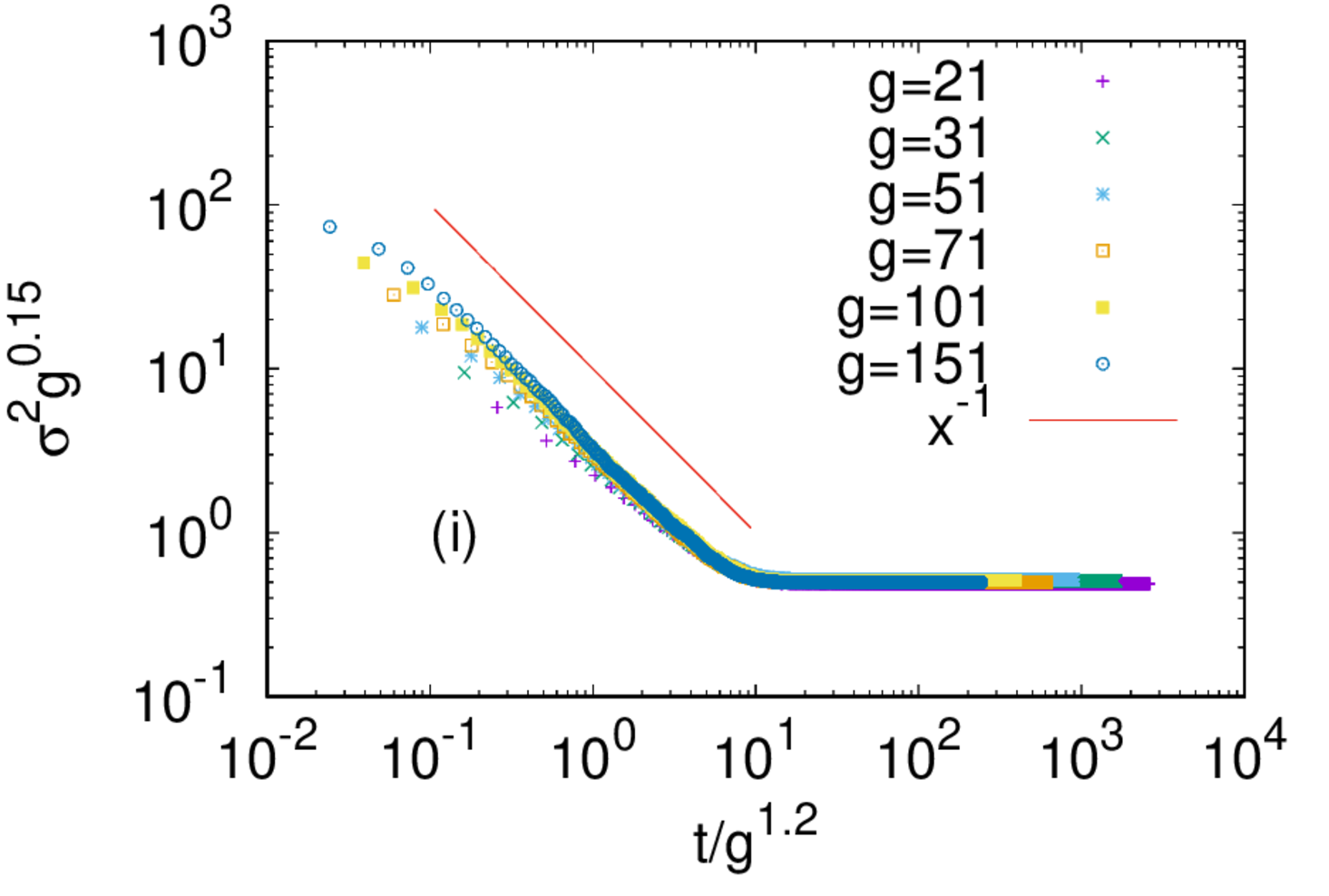}
\includegraphics[width=5.5cm]{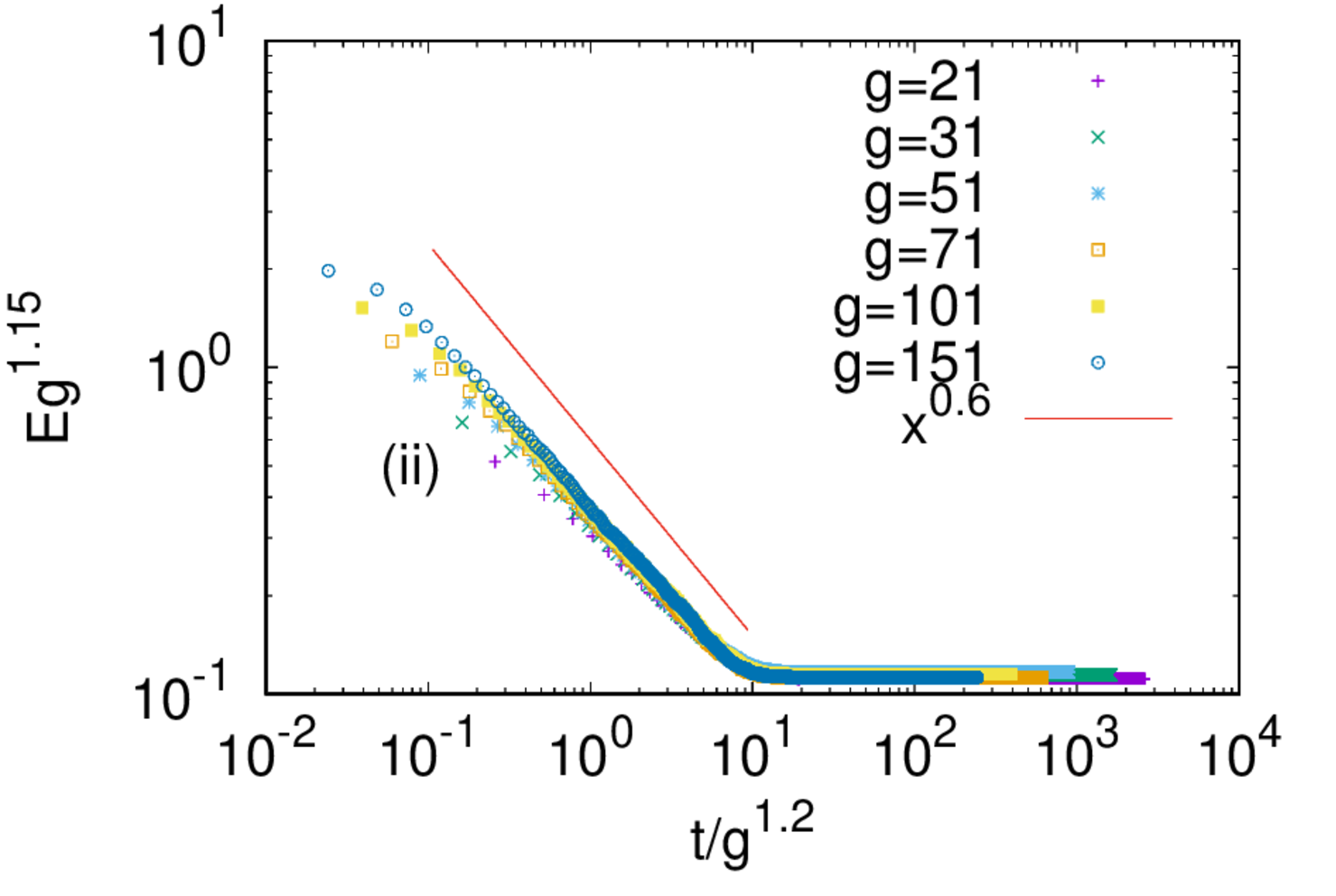}
\includegraphics[width=5.5cm]{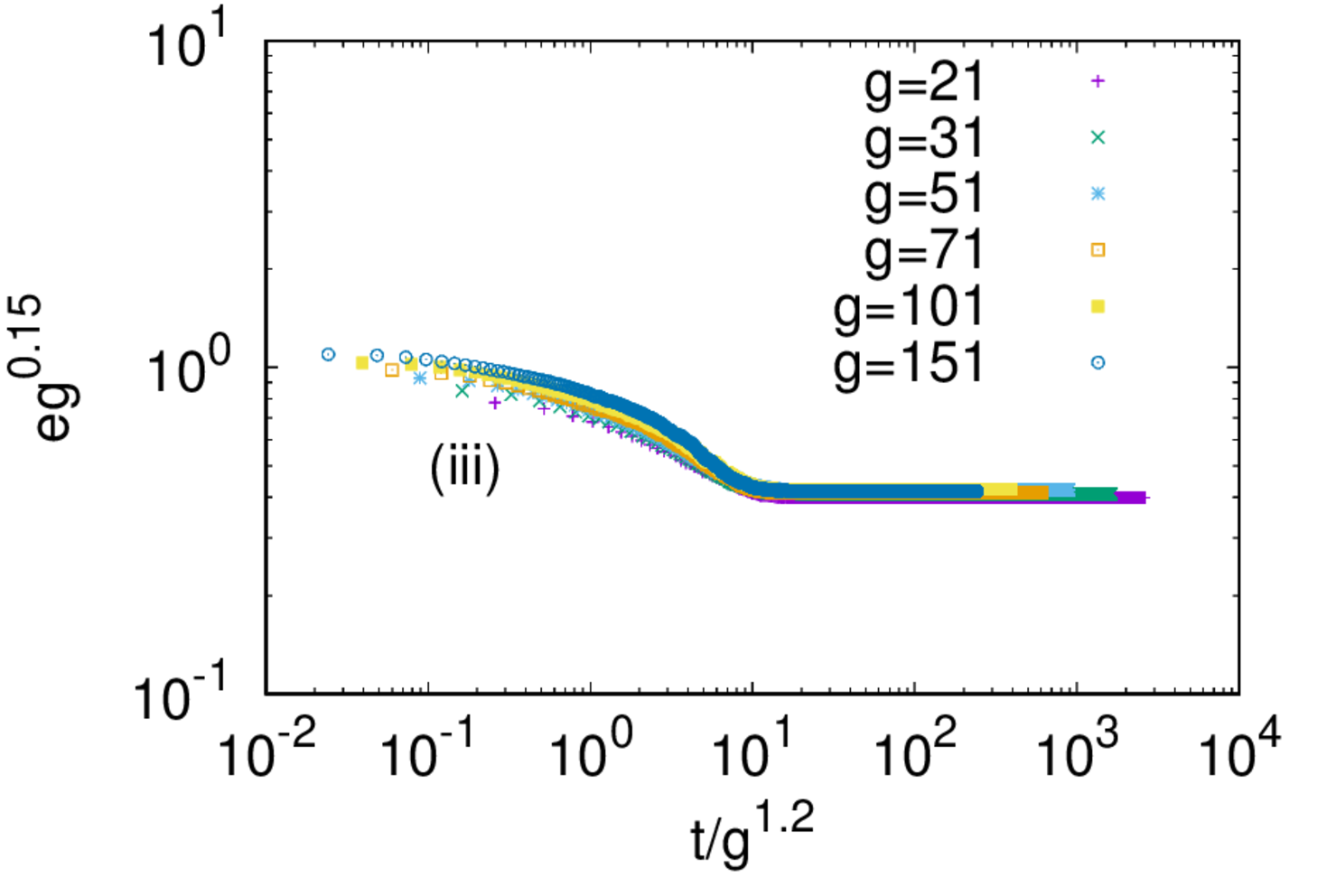}
\caption{The scaling properties of the three measures of population fluctuation $\sigma^2(g,t)$ (Eq. (\ref{eq_rms})), $E(g,t)$ (Eq. (\ref{eq_E})) and $e(g,t)$ (Eq. (\ref{eq_e})) are shown for strategy C (Eq. (\ref{eq_stratC})). It is worth recalling that a complete random choice $\sigma^2\sim g$, while in this strategy $\sigma^2(t\to\infty)\sim g^{-0.15}$, within our numerical accuracy. On the other hand, the time required to reach this saturation value scales as $g^{1.15}$, as opposed to $g^0$ in the random strategy.}
\label{fig_5}
\end{figure*}

\subsection{Strategies}
We now move on to describe the strategies that could be followed by the agents in order to reduce volatility and maximize the chance to stay in the minority between each agent's two choices.

 {\bf Strategy A}: If the total population is $N$, then the ideal distribution would be (the nearest integer to) $g=N/D$ agents in each choice, on average over time. Therefore, in line with the stochastic strategy developed in ref. \cite{sasi,utl}, one can assume a switching probability for the $i$-th agent as
\begin{equation}
p_i^A(t)=\frac{g-n_{x_i}(t)}{2n_{x_i}(t)},
\label{eq_stratA}
\end{equation}
where $x_i(t)$ is the location of the $i$-th agent at time $t$, and $n_{x_i}(t)$ is the population at that choice at that time. 
This is expected to drive the population in each choice towards the average. The switching probability is only activated when the population in a choice is above the expected global average $g=N/D$. This strategy does not take the minority consideration between each choices into account i.e., an agent sitting at the minority of their two choices can also switch.

 {\bf Strategy B}: In this strategy, the agents are supplied the additional information of the population on their alternate choice at the same instance of time (same information level as that in ref. \cite{sasi}). An agent may switch, with a probability 
\begin{equation}
p_i^B(t)=\frac{n_{x_i}(t)-n_{y_i}(t)}{2n_{x_i}(t)}, 
\label{eq_stratB}
\end{equation}
where $y_i(t)$ is the alternate for that particular agent at that instance in time. Here, the switching only happens if the agent is in the majority of their two alternative i.e., $n_{x_i}(t)>n_{y_i}(t)$, otherwise, $p_i^B=0$.  

 {\bf Strategy C}: Finally, we restrict the information regarding population at the alternate location in the following way: the agents only retain the memory of the population at the last instance of their stay at their alternate location, and only switches with a probability if the population in their present location is higher than their memory of the population in their alternate location. Of course, the timing of their last visits to the alternate could be different for each agent. The switching probability is 
\begin{equation}
p_i^C(t)=\frac{n_{x_i}(t)-n_{y_i}(t^{\prime})}{2n_{x_i}(t)},
\label{eq_stratC}
\end{equation}
where $t^{\prime}<t$ always. The agents do not switch if their memory of the alternate location is worse (having higher crowd) than their present experience (or crowd).

\section{Results}
Here we discuss the numerical simulation results corresponding to each of the strategies and quantification of the population fluctuation mentioned before. 

\begin{figure*}
\includegraphics[width=7.5cm]{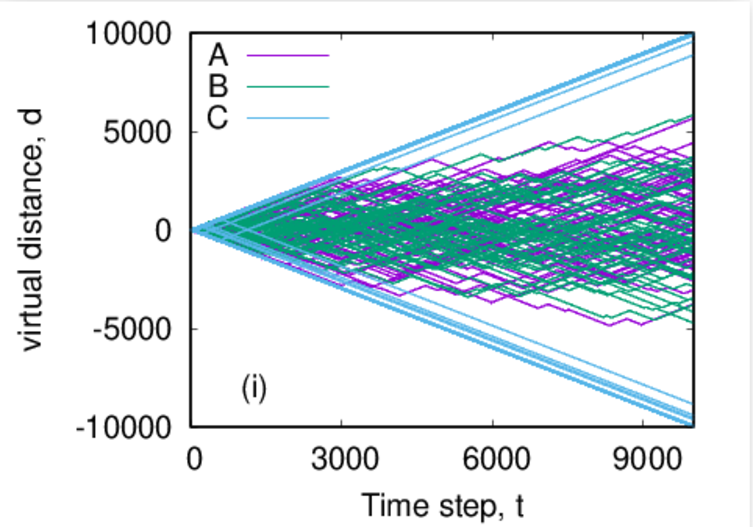}
\includegraphics[width=7.5cm]{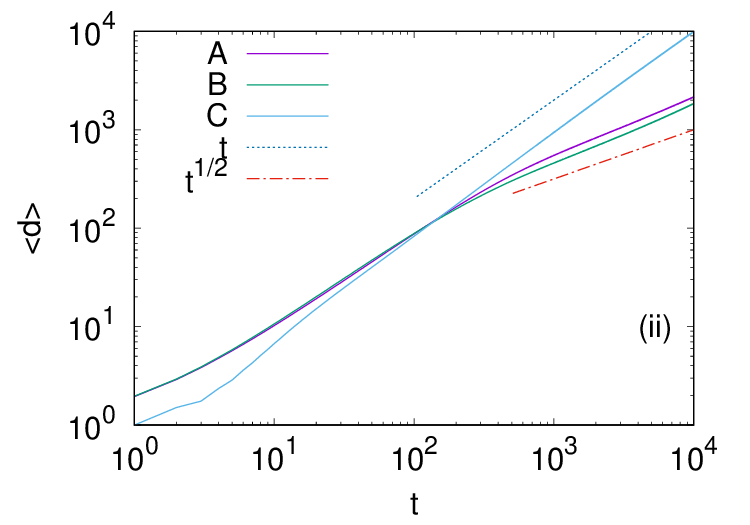}
\caption{(i) The paths of the virtual walkers, defined in Eq. (\ref{map}) are shown for the three strategies. The paths for strategy A and strategy B seem diffusive, while that for strategy C is ballistic in the long time limit. The simulations are for $g=51$. (ii) The average distance traveled by the walkers indicate that in the long time it scales are $t^{1/2}$ for strategies A and B, and linearly in $t$ for strategy C. The short time behavior indicate an opposite trend. The ballistic paths for strategy C is due to the freezing of dynamics.}
\label{fig_6}
\end{figure*}

Unless otherwise specified, we keep $D=101$, and vary $g (=N/D)$. In Fig. \ref{fig_2}, we show the variance (see Eq. (\ref{eq_rms})) for the three strategies mentioned in the last section. This is a standard measure for the MG to quantify efficiency of a strategy. For strategy A, The time variation suggests that a steady state is reached very quickly, with very little to no dependence on the population size. There is some dependence of the variance at the saturation level on the system size, and it is increasing very slightly. For strategy B, however, there is no detectable system size dependence of either the saturation time or the saturation variance. More over, both the saturation time and saturated variance are smaller than that obtained for strategy A. This strategy is closest to what was done for stochastic strategy MG, and there it was shown that the saturation time has $\tau\propto\log \log N$ variation. In the range of $g$ studied here, such a dependence would not be noticeable. Finally, for strategy C, the saturation variance is lower than what could be obtained for either of the earlier two strategies, at least for the higher values of $g$. However, the saturation time is significantly longer, as the variance decays in a power-law with time. We will return to the question of the system size scaling of the saturation time for strategy C later on. But an important point to mention here is that at and above the saturation time, the dynamics freeze for this strategy. This means that the agents get frozen in their choices after the variance reaches a low value. However, it is clear that such a configuration may not be the best possible configuration for the system i.e., some agents get frozen in the majority just because the memory of their alternate choice is worse.  

\begin{figure*}
\includegraphics[width=8.5cm]{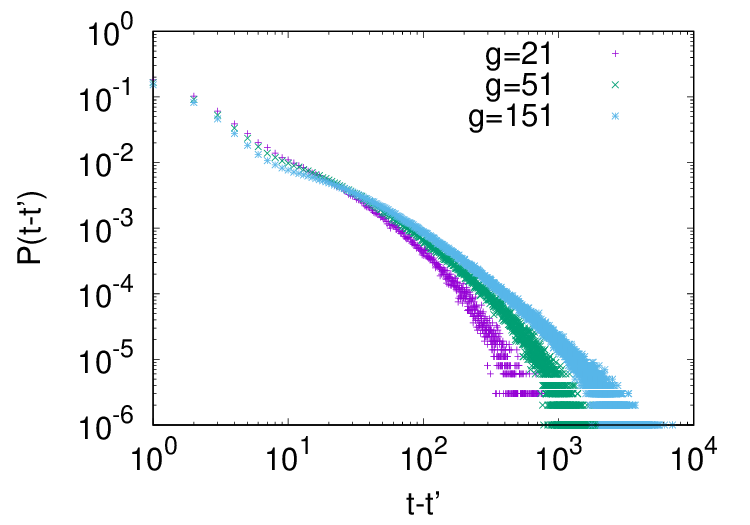}
\caption{The distribution of the time lag $t-t^{\prime}$ are shown for strategy C with different values of $g$ sampled only for the agents who are about a make a switch is strategy C. The wide distributions confirm that for many agents the time elapsed between their successive visits to the alternate is large, leading to an emergent crowd avoidance in relatively emptier choices, which in turn leads to a more uniform distribution across the choices and hence a lower variance. On the other hand, the large time lags lead to outdated information for the agents, which leads to freezing of dynamics.}
\label{fig_7}
\end{figure*}

We also measure the other two quantities denoting fluctuation (see Eq. (\ref{eq_E} and (\ref{eq_e}))). These are shown in Fig. \ref{fig_3} and Fig. \ref{fig_4} respectively. Of course, the time to reach the steady state has the same variation as that for the variance. However, we see that for each of the strategies $E(t\to\infty)$ decreases significantly with $g$. But for strategy C it reaches the lowest value. On the other hand, for $e(t\to\infty)$ there is little to no system size dependence for the strategies A and B. However, there is a decrease with system size for the strategy C. Also, the minimum value for this measure is obtained by following strategy C. These results show that in all the three measures for population fluctuation defined before, strategy C performs the best, notwithstanding the freezing behavior there. 

As strategy C is the only strategy where a significant system size dependence is noted in all three measures, we go on to study its finite size scaling. Fig. \ref{fig_5} depicts the finite size scaling behavior of the three measures for strategy C. It is clear that the variance follow a finite size scaling of the form
\begin{equation}
\sigma^2_C=g^{-\theta_1}F\left(t/g^{\theta_2}\right),
\end{equation}
with $\theta_1\approx 0.15$, $\theta_2\approx 1.2$ and $F(x)\sim 1/x$ when $x<<1$ and $F(x)\sim const.$ for $x>>1$. We have kept the subscript C to the variance to underline that this is done only for strategy C. Similar scaling behavior for $e_C$ and $E_C$ are also shown in Fig. \ref{fig_5}, with different exponent values and functional form of the scaling function.

Therefore, it is generally seen that for strategy C, a larger system size leads to a relatively better saturation configuration in terms of the reduced fluctuation. 

Then, we look at the virtual walker, as defined in Eq. (\ref{map}). Fig. \ref{fig_6} shows the paths of 50 such walkers in each of the three strategies. There is no qualitative difference between strategies A and B, and in both cases the walks look diffusive. However, for strategy C, the walk is clearly ballistic at least at long time. We have also measured the average distances traveled by such walkers, which show the diffusive ($\langle d\rangle \sim t^{1/2}$) scaling for the strategies A and B, and ballistic ($\langle d\rangle \sim t$) scaling for strategy C. The early time behavior in both cases seem to suggest an opposite trend. The long term ballistic trend of the virtual walker in strategy C suggests that the agents memory of the alternate choice becomes vastly outdated and therefore they can end up being stuck in their current location. Indeed, Fig. \ref{fig_7} shows the distribution of the time lag between the time ($t$) at which an agent switches options following strategy C and the time ($t^{\prime}$) of the memory of the alternate location they used while switching. It shows a fat-tailed distribution, which is not a power law. Nevertheless, this outdated memory, generally representing an early dynamics of uneven population distribution,  keeps the agents stuck in their current choice. Also, prior to being stuck, this time-lag also helps a more uniform population distribution, by creating crowd avoidance from multiple locations to a relatively less populated choice.

\section{Discussion and conclusions}
The Minority Game (MG) is a well-studied model of competitive resource allocation among independent agents through repeated binary choices and forms a paradigmatic example of emergent cooperation driven by shared strategies. A stochastic strategy version of the model has been the most successful in minimizing the fluctuation between the choices, even though the dynamics freezes at that point. 

Here we have studied the parallel MG, which is essentially a set of coupled MG with overlapping choices and its population. Even though there are $D$ choices in total, for a given agent there are only two allowed choices. This means that for any two choices, there are a set of agents for whom the situation is just like the MG. However, in each choice, there are other agents present as well, who can have their alternate choices anywhere else among the remaining $D-2$ options. Finding the optimal distribution of the population -- uniform for all choices -- in a non-dictated manner seems harder than the standard MG, at least with the stochastic strategies that work for the standard MG. This is because, the minimization of population fluctuation between any two choices are not solely dependent upon the dynamics of the agents moving between those two choices. But due to the coupling introduced through the other agents who have their alternate choices elsewhere, the entire system of $D$ choices are coupled. Finding the most effective distribution of population is, therefore, a combinatorial optimization problem. Unlike with the stochastic strategy MG, here a residual variance remains in all strategies that were tried. In that way, it is reminiscent of frustrated disordered systems, such as spin glasses. Indeed, the best strategy in terms of the lowest variance (and similar other measures) for PMG, does leave the system in a frozen state, even though that is clearly not the state with the minimum possible variance for that configuration. 

The strategy that works best (given by Eq. (\ref{eq_stratC})) for the PMG is the one where the agents only work with the information regarding the population at their current location in the present time and the information regarding the population of their alternate location at the time of their last stay there. This time lag in the available information naturally prevents overcrowding at a relatively less populated location at a particular time. This overcrowding is presumably the reason why in the other two strategies, where decisions are made with instantaneous information only, variance is higher. However, the cost for the strategy C is the longer time taken to reach the saturation in the variance. Additionally, as the virtual walk picture reveals (see Fig. \ref{fig_6}), the dynamics get frozen at that minimum variance state. Notwithstanding these two issues, the finite size scaling of the variance and other related measures for this strategy suggest an increasingly better solution (more uniform population distribution) with increasing system size. 

In conclusion, we have studied various stochastic strategies for reducing fluctuation in population in the parallel Minority Game, which is essentially a coupled set of standard Minority Game. The overlap of choices and population give rise to frustration that can only partially be reduced with a help of a time-delayed stochastic switching strategy of the agents. 

\section*{Acknowledgements}

The authors are grateful to Bikas K. Chakrabarti for discussions and comments on the manuscript. The simulations were performed in HPCC Chandrama in SRM University - AP.

\section*{Author contributions statement}

S.B. conceptualized the problem, A.R.V. performed the simulations, both the authors analysed the results. S.B. wrote the manuscript. Both the authors reviewed the manuscript.





\end{document}